\RequirePackage{fix-cm}
\documentclass[onecollarge]{svjour3}
\smartqed  
\usepackage{xspace}
\usepackage{graphicx}
\usepackage{amssymb,amsbsy,amsmath}
\usepackage{lineno}
\usepackage{hyperref}
\usepackage{color}
\modulolinenumbers[5]
\newcommand{\nc}[1]{\newcommand #1}
\newcommand{\rnc}[1]{\renewcommand #1}
\nc{\red}[1]{#1}
\nc{\myquote}[1]{`#1'}
\nc{\x}[1]{\mbox{#1}}
\rnc{\matrix}[2]{\left[\!\!\begin{array}{#1} #2\end{array}\!\!\right]}
\rnc{\vector}[1]{\matrix{c}{#1}}
\nc{\suml}[2]{\sum \limits_{#1}^{#2}}
\nc{\real}[1]{\Re\lbrace #1 \rbrace}
\nc{\imag}[1]{\Im\lbrace #1 \rbrace}
\nc{\conj}{\overline}
\nc{\order}[1]{\mathcal O\left(#1\right)}
\nc{\tsclone}{\tau}
\nc{\tscltwo}{\eta}
\nc{\sigone}{\sigma_1}
\nc{\sigtwo}{\sigma_2}
\nc{\fuh}{F_u}
\nc{\fvh}{F_v}
\nc{\fone}{F}
\nc{\cc}{\mathrm{c.c.}}
\nc{\g}[1]{\x{$#1$}}
\nc{\e}[2]{\begin{equation} #1 \label {eq:#2} \end{equation}}
\nc{\eal}[2]{\begin{equation} \begin{aligned} #1 \label {eq:#2} \end{aligned} \end{equation}}
\nc{\ea}[2]{\begin{eqnarray}
#1 \label {eq:#2}
\end{eqnarray}}
\nc{\inv}{^{-1}}
\nc{\tra}{^{\mathrm T}}
\nc{\herm}{^{\mathrm H}}
\nc{\fabstand}{\,}
\nc{\fp}{\fabstand.}
\nc{\fk}{\fabstand,}
\nc{\mm}[1]{\mathbf{#1}}
\nc{\mms}[1]{\boldsymbol{#1}}
\nc{\ie}{i.\,e.\xspace}
\nc{\eg}{e.\,g.\xspace}
\nc{\cf}{cf.\xspace}
\nc{\dd}{{\mathrm{d}}}
\nc{\ii}{{\mathrm{i}}}
\nc{\jj}{\ii}
\nc{\ee}{{\mathrm{e}}}
\nc{\ommod}{{\omega_0}}
\nc{\gam}{\gamma}
\nc{\del}{\delta}
\nc{\phaselag}{\Delta\beta}
\nc{\dmod}{D}
\nc{\omone}{{\omega_1}}
\nc{\omtwo}{{\omega_2}}
\nc{\omonesq}{{\omega_1^2}}
\nc{\omtwosq}{{\omega_2^2}}
\nc{\kc}{k_{\mathrm c}}
\nc{\fref}[1]{\x{Fig.~\ref{fig:#1}}}
\nc{\eref}[1]{\x{Eq.~(\ref{eq:#1})}}
\nc{\erefs}[1]{\x{Eqs.~(\ref{eq:#1})}}
\nc{\erefo}[1]{(\ref{eq:#1})}
\nc{\tref}[1]{\x{Tab.~\ref{tab:#1}}}
\nc{\sref}[1]{\x{Sec.~\ref{sec:#1}}}
\nc{\srefo}[1]{\ref{sec:#1}}
\nc{\srefs}[1]{\x{Sec.~\ref{sec:#1}}}
\nc{\ssref}[1]{\x{Subsec.~\ref{sec:#1}}}
\nc{\aref}[1]{\x{Appendix~\ref{asec:#1}}}
\nc{\fig}[3][tbh]{
\begin{figure}[#1]
\centering
\includegraphics{figures/#2}
\caption{#3\label{fig:#2}}
\end{figure}}
\nc{\figc}[3][tbh]{
\begin{figure}[#1]
\centering
\includegraphics[width=0.5\columnwidth]{figures/#2}
\caption{#3\label{fig:#2}}
\end{figure}}
\nc{\figw}[3][tbh]{
\begin{figure*}[#1]
\centering
\includegraphics[width=1.0\textwidth]{figures/#2}
\caption{#3\label{fig:#2}}
\end{figure*}}
\journalname{Archive of Applied Mechanics}
\begin{document}

\title{Global Complexity Effects due to Local Damping in a Nonlinear System in 1:3 Internal Resonance
}

\titlerunning{Global Complexity due to Local Damping near 1:3 Internal Resonance}        

\author{Malte Krack         \and
        Lawrence A. Bergman \and
        Alexander F. Vakakis 
}


\institute{M. Krack \at
              Institute of Dynamics and Vibration Research,
              Leibniz Universit\"at Hannover,
              Appelstr. 11, 30167 Hannover, Germany\\
              \email{krack@ids.uni-hannover.de}           
           \and
           L. A. Bergman \at
              Department of Aerospace Engineering, University of Illinois at Urbana-Champaign, 104 S. Wright Street, Urbana, IL 61801, USA
           \and
           A. F. Vakakis \at
              Department of Mechanical Science and Engineering, University of Illinois at Urbana-Champaign, 1206 W. Green Street, Urbana, IL 61801, USA
}

\date{Received: date / Accepted: date}

\maketitle

\begin{abstract}
It is well-known that nonlinearity may lead to localization effects and coupling of internally resonant modes. However, research focused primarily on conservative systems commonly assumes that the near-resonant forced response closely follows the autonomous dynamics. Our results for even a simple system of two coupled oscillators with a cubic spring clearly contradict this common belief. We demonstrate analytically and numerically global effects of a weak local damping source
in a harmonically forced nonlinear system under condition of 1:3 internal resonance: The global motion becomes asynchronous, \ie, mode complexity is introduced with a non-trivial phase difference between the modal oscillations. In particular, we show that a maximum mode complexity with a phase difference of $90^\circ$ is attained in a multi-harmonic sense. This corresponds to a transition from generalized standing to traveling waves in the system's modal space. We further demonstrate that the localization is crucially affected by the system's damping. Finally, we propose an extension of the definition of mode complexity and mode localization to nonlinear quasi-periodic motions, and illustrate their application to a quasi-periodic regime in the forced response.

\end{abstract}

\section{Introduction\label{sec:intro}}
The behavior of mechanical systems can in general be described by nonlinear differential equations. The nonlinearity may have various physical sources, \eg, large deflections, hysteretic material behavior or contact interactions. In the typical operation regime of many state-of-the-art engineering applications, nonlinear effects cannot be neglected. In fact, they might even be intentionally utilized, as for instance in the case of nonlinear energy sinks \cite{vaka2008b}. The design of such systems requires a thorough understanding of the underlying dynamics and appropriate simulation and experimental testing methods. In this context, certain critical situations are often of particular concern. In the presence of external forcing, the phenomenon of resonance coincidence is probably the most common one. Since nonlinearity distorts the temporal and spatial vibration spectra, it can lead to the coupling of otherwise distinct and widely spaced modes. This modal interaction phenomenon is particularly interesting, as it can lead to entirely unexpected behavior from a linear dynamics perspective.\\
Nonlinear modal interactions have been the subject of many theoretical investigations, and experimental studies \cite{Bux.1986,bert1998,noel2014}. Frequently reported effects on the steady-state forced response are
\begin{itemize}
\item suppression of the linear resonance peak, while one or more indirectly excited modes feature significant responses,
\item highly nonlinear responses including bent amplitude-frequency curves leading to hysteresis loops, and
\item quasi-periodic regimes associated with possibly strong beating phenomena, as well as
\item vibration localization.
\end{itemize}
The completeness of the suppression of linear resonance peaks and the response levels of indirectly excited modes depends on damping. Large damping tends to diminish the response of the indirectly excited modes. Moreover, nonlinear interaction is likely to occur if the natural frequencies of the associated modes are low-integer multiples of each other giving rise to internal resonance.\\
Under typical operating conditions, the dynamic behavior of most engineering structures can be characterized as dissipative, and often a permanent source of excitation is present. In spite of this, most theoretical investigations of nonlinear modal interactions focus on symmetric, \ie, non-gyroscopic and non-circulatory, conservative systems. This approach is useful as long as damping and forcing only play a minor role in the sense that they primarily govern the response level; however the spatiotemporal vibration content closely follows the dynamics of the underlying conservative system. This appears to be justified if the damping is relatively weak, of modal type\footnote{In this article, a linearized system with symmetric structural matrices is considered to feature modal (non-modal) damping if its modal deflection shapes are (not) identical to the ones in absence of damping. In the case of modal damping, the damping matrix is diagonal in the modal space. A special but common case of modal damping is proportional, or Rayleigh, damping, where the viscous damping matrix is a linear combination of the mass and the stiffness matrix.} and more or less evenly distributed among the modes. This type of damping is often assumed for global dissipation mechanisms such as  viscous damping by an ambient medium or slightly hysteretic material behavior. In contrast, mechanical joints or attachments subject to dry friction may introduce a rather local dissipation source which leads to a highly non-modal damping distribution in the system.\\
Even in the linear case, local damping may have a crucial effect on the resulting dynamics. In particular, it can give rise to mode complexity, \ie, a non-trivial phase lag between the oscillations of different material points of a system \cite{DasGupta.2007,Blanchard.2015a}. Depending on the system parameters and the degree of asyncronicity of the oscillation, the dynamics may change completely from standing waves to a traveling waves \cite{Blanchard.2015a,Blanchard.2015b}. It should be noted that non-unison vibrations can also occur in undamped nonlinear systems. Examples are the traveling waves in cyclic periodic structures \cite{vaka1992b}, and the whirling motions of shallow cables \cite{Nielsen.2003}. In the case of whirl, a phase lag of about $90^\circ$ between the lateral deflection coordinates may occur so that the material points undergo an elliptic motion. This phenomenon is not limited to in-plane vibrations, but has recently been observed in a chain of two oscillators in $1:1$ internal resonance condition \cite{Hill.2015}.\\
To the authors' knowledge, mode complexity in the neighborhood of an internal resonance has not yet been investigated. Our objective is to understand the emergence of mode complexity and gain knowledge on the effect of local damping and harmonic forcing on the nonlinear interactions between two modes in internal resonance. We therefore analyze a two-degree-of-freedom oscillator whose natural frequencies are near a ratio of $1:3$. The system is described in \sref{system}. We are mostly interested steady-state vibrations, so we study the vibration response to harmonic external forcing in the neighborhood of the first linear natural frequency. Analytical investigations using multiple scales are reported in \sref{msa}. The results are illustrated and discussed in \sref{results}. This section also includes a validation of the asymptotic results with numerical integrations. Moreover, the quasi-periodic regime in the neighborhood of the internal resonance is analyzed, and a generalization of mode complexity and mode localization for this case is proposed. The main findings are summarized in \sref{conclusions}.

\section{Model description\label{sec:system}}
\figc[b!]{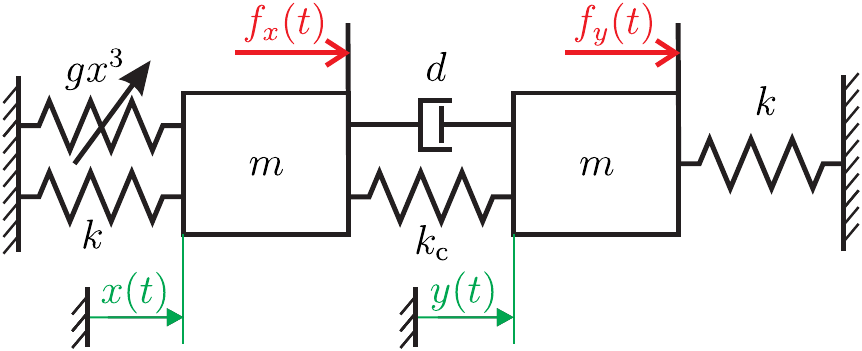}{System of two coupled oscillators}
Consider the system of two coupled oscillators depicted in \fref{two_dof_oscillator}. The underlying linear system is symmetric, while the cubic spring that grounds the left oscillator perturbs the symmetry in the nonlinear setting. It should be noted that a single viscous damper is placed between the oscillators with the intention to introduce a local dissipation source, and, hence render the damping distribution non-modal.\\
Considering the nonlinear system, in terms of the physical coordinates $x(t)$, $y(t)$, the dynamics is governed by a set of two second-order differential equations,
\ea{m\ddot x + d\left(\dot x-\dot y\right) + \kc\left(x - y\right) + kx& + gx^3 =& f_x(t)\fk\label{eq:eqm_xya}\\
m\ddot y + d\left(\dot y-\dot x\right) + \kc\left(y - x\right) + ky& =& f_y(t)\fp}{eqm_xy}
Herein, overdot denotes derivative with respect to time. For brevity, the time dependence of coordinates, \eg, \g{x(t), y(t)} is not explicitly denoted here and in the following.\\
When the nonlinearity is absent, the underlying linear system possesses two widely spaced modes, and modal interaction is possible. In that case, assuming that the damping is small, its effects are only confined in the neighborhoods of the two linear resonances and no notable mode complexity is possible. The linearized modes can be identified as symmetric in-phase and out-of-phase. For further analysis, it is convenient to introduce the coordinate transform $\left[x\,\,y\right]\tra = \left[1\,\,1\right]\tra u + \left[1\,\,-1\right]\tra v=\mm T\left[u\,\,v\right]\tra$ so the problem is expressed in (linear) modal form,
\ea{\red{\ddot u}& \red{+} &\red{\omonesq u + \frac{\gam}{2}\left(u+v\right)^3 = f_u(t)\fk}\label{eq:eqm_uva}\\
\red{\ddot v}& \red{+ 2\del\dot v +} &\red{\omtwosq v + \frac{\gam}{2}\left(u+v\right)^3 = f_v(t)\fk}}{eqm_uv}
with \g{\left[f_u\,\,f_v\right]\tra = \frac{1}{2m}\mm T\tra\left[f_x\,\,f_y\right]\tra}. The first linearized (in-phase) mode is undamped and has a natural frequency of \g{\omonesq=k/m}, which is independent of the coupling parameters \g{\kc} and \g{d}. In contrast, the second linearized (out-of-phase) mode is damped, and its natural frequency depends on \g{\kc} with \g{\omtwosq=(k+2\kc)/m}. \red{The damping parameter $\del$ is related to $d$ via $\del=d/m$, and the nonlinear stiffness parameter $\gam$ is defined as $\gam = g/m$.} For \g{\kc=4k}, the modes are in a $1:3$ internal resonance situation with \g{\omtwo=3\omone}.\\
In this paper, the vibration response of the system of coupled oscillators subject to harmonic forcing is investigated. To this end, general harmonic excitations acting on the two linear modal coordinates are considered as,
\e{f_u(t) = \fuh\ee^{\ii\Omega t}+\cc \fk\quad f_v(t) = \fvh\ee^{\ii\Omega t}+\cc\fp}{harmonic_forcing}
Herein, the amplitudes $\fuh$ and $\fvh$ are in general complex-valued quantities to account for a possible phase lag of the forcing, and \g{\cc} denotes the complex conjugate of the preceding terms. Of particular interest will be harmonic excitations with frequency close to the in-phase (first) linearized mode, that is, $\Omega\approx\omone$.

\section{Asymptotic analysis\label{sec:msa}} 
\subsection{Analytical investigation of the slow flow}
As the exact solution of \erefs{eqm_uva}-\erefo{eqm_uv} cannot be expressed in closed form, we resort to the method of multiple scales
. To this end, we assume that nonlinearity, forcing, and damping are weak and of $\order{\epsilon}$, where $\epsilon\ll 1$ is a small parameter. Hence, we rescale \erefs{eqm_uva}-\erefo{eqm_uv} with $\gam=\epsilon \gam^*$, $\del=\epsilon \del^*$, $f_u=\epsilon f_u^*$, $f_v=\epsilon f_v^*$. In addition, small external and internal detuning parameters $\sigone$ and $\sigtwo$ are assumed, defined as
\e{\Omega = \omone + \epsilon\sigone \fk\quad \omtwo = 3\omone + \epsilon\sigtwo\fp}{detuning}
Thus, the frequency around the first linear natural frequency is externally forced, and the system is close to a $1:3$ internal resonance condition.\\
Here, we only present the results here and refer to \aref{derivation_slowflow} for details on the mathematical developments. The approximate solution of \erefo{eqm_uva},\erefo{eqm_uv} takes the form
\ea{u(t) \approx a_1\ee^{\ii\left(\Omega t+\beta_1\right)} + \cc\fk\,\, v(t) \approx a_2\ee^{3\ii\left(\Omega t+\beta_2\right)} + \cc\fp}{uv_solution}
The slow flow of the forced response in terms of the real-valued quantities $a_1$, $a_2$, $\beta_1$ and $\beta_2$ is governed by a set of four first-order differential equations,
\ea{
2\omone a_1^\prime =& -\fone\sin\beta_1 - &\frac{3\gam}{2} a_1^2a_2\sin\phaselag\fk\label{eq:slowflowa}\\
2\omone a_1\beta_1^\prime =& -\fone\cos\beta_1 + &\frac{3\gam}{2}\left[a_1^3+2a_1a_2^2+a_1^2a_2\cos\phaselag\right] \nonumber\\
 && - 2\omone\sigone a_1\fk\label{eq:slowflowb}\\
6\omone a_2^\prime = &&\frac{\gam}{2}a_1^3\sin\phaselag-6\omone da_2\fk\label{eq:slowflowc}\\
18\omone a_2\beta_2^\prime = &&\frac{\gam}{2}\left[3a_2^3+6a_1^2a_2+a_1^3\cos\phaselag\right]\nonumber\\
 && + 6\omone\left(\sigtwo-3\sigone\right)a_2\fk
}{slowflow}
with the abbreviation $\phaselag=3\left(\beta_2-\beta_1\right)$. As it was noted during the derivation, the assumed fundamental harmonic forcing of the second mode as defined in \eref{harmonic_forcing} has only higher order effects. Without loss of generality, the complex amplitude $\fuh$ was therefore replaced by a real-valued magnitude $\fone$. It can be easily seen from \erefs{slowflowa}-\erefo{slowflow} that only the first mode is directly excited, while only the second mode is directly damped. However, due to the $1:3$ internal resonance it will be shown that energy can be transferred from the lower-frequency to the higher-frequency mode. The character of this modal interaction is essentially nonlinear.

\subsection{Analytical investigation of the steady state responses}
The periodic steady-state response of system \erefo{eqm_uva},\erefo{eqm_uv} can be approximated by determining the stationary points of \erefs{slowflowa}-\erefo{slowflow}, \ie, by setting all time derivatives to zero and solving the resulting set of algebraic equations. \red{In general, the (multitude of possible) solutions and their stability can be analytically determined from \erefs{slowflowa},\erefo{slowflow}. However,} we are primarily interested in mode localization and mode complexity, so we focus on studying the ratio between the amplitudes $a_1$ and $a_2$, as well as the phase lag $\phaselag$. 
These quantities can be obtained considering only \erefs{slowflowc}-\erefo{slowflow}, while the remaining equations \erefo{slowflowa}-\erefo{slowflowb} merely govern the appropriate force magnitude $\fone$ and phase $\beta_1$ consistent with these steady-state periodic responses. Hence, it is sufficient to focus on \erefs{slowflowc}-\erefo{slowflow} in the following.\\
For convenience, we introduce yet another transformation of the real steady-state amplitudes:
\e{a_1=\rho\cos\psi \fk\quad a_2=\rho\sin\psi\fp}{rhotrafo}
The results obtained for a certain $\rho$ can be interpreted as forced response with controlled amplitude level $\rho^2=a_1^2+a_2^2$, as opposed to the usual controlled excitation level $\fone$. The angle $\psi$ defines the amplitude ratio between $a_1$ and $a_2$ and therefore the localization of the steady-state response in either of the two nonlinearly interacting modes in $1:3$ internal resonance. Substituting \eref{rhotrafo} into \erefs{slowflowc}-\erefo{slowflow} yields
\ea{\sin\phaselag &=& 12\underbrace{\frac{\omone \del}{\gam\rho^2}}_{\del^*}\frac{\sin\psi}{\cos^3\psi}\fk\label{eq:sincosdeltaa}\\
\cos\phaselag &=& 12\underbrace{\frac{\omone(3\sigone-\sigtwo)}{\gam\rho^2}}_{\sigma^*}\frac{\sin\psi}{\cos^3\psi} \nonumber\\
&&\quad\quad - 3\frac{\sin\psi}{\cos\psi}\left(2+\frac{\sin^2\psi}{\cos^2\psi}\right)\fk}{sincosdelta}
Using the trigonometric identity $\sin^2\phaselag+\cos^2\phaselag=1$, these equations can be combined into a single bi-cubic equation in $\sin\psi$,
\ea{\nonumber &10{\sin^6\psi} + \left(72\sigma^*-39\right){\sin^4\psi} + \\
&\quad\quad\quad\left(144\left({\sigma^*}^2+{\del^*}^2\right)-144{\sigma^*}+39\right){\sin^2\psi} = 1\fk}{bicubic}
whose exact roots are well-known, but not given here for the sake of brevity. Once $\psi$ is determined, \erefs{sincosdeltaa}-\erefo{sincosdelta} can be evaluated to determine $\phaselag$. Note that these quantities, and therefore the mode complexity and localization only depend on the two unitless parameters $\del^*$ and $\sigma^*$. These parameters can be interpreted as normalized damping and normalized detuning parameters, respectively. It immediately follows from the definition of $\sigma^*$ that external $\sigone$ and internal $\sigtwo$ detuning have similar effects. This generally resembles the findings reported by Bux and Roberts \cite{Bux.1986}: They studied the forced response of two coupled beams with geometric nonlinearity leading to quadratic terms in the equation of motion. They investigated an internal resonance situation and found that the response ratio depends solely on the damping and the frequency ratio of the participating modes.

\subsection{Maximum mode complexity}
To avoid spurious solutions generated by the periodicity of the trigonometric functions, \red{these} ranges are considered for the respective quantities in the following: $a_1\geq 0$, $\rho\geq 0$ and $\psi\in[-\pi/2\,,\,\,\pi/2)$. Note that these ranges readily span the entire solution space so that the subsequent developments hold without loss of generality.\\
Presuming that there exists a point with $\phaselag=\pi/2$, \eref{sincosdeltaa} can be solved for $\psi$. This leads to the following depressed bi-cubic in $\cos\psi$,
\e{\cos^6\psi + 144{\del^*}^2\cos^2\psi - 144{\del^*}^2 = 0\fp}{c2crit_eqn}
It can be verified that the discriminant is negative, and this equation has exactly one real solution
\e{\cos^2\psi = \del^{**}\sinh\left[\frac{\sinh^{-1}\frac{3}{\del^{**}}}{3}\right]\fk}{c2crit_sol}
with $\del^{**} = 8\sqrt{3}\del^*$. The existence of one real solution confirms the presumption that there is always a point of maximum mode complexity. At this point, the localization only depends on $\del^*$ which follows from \eref{c2crit_sol}. It can be verified that the function on the right-hand side of \eref{c2crit_sol} is strictly increasing with $\del^*$. Thus, $\cos\psi$ increases with $\del^*$. Considering \eref{rhotrafo}, it follows that the motion localizes in the first mode for increasing $\del^*$. In other words, an increasing damping reduces the mode mixing and damps away the super-harmonic resonance.\\
In the case of $\phaselag=\pi/2$, the function $\sin\phaselag$ assumes a maximum. Further, the function $\sin\psi/\cos^3\psi$ on the right-hand side of \eref{sincosdeltaa} is strictly increasing in the considered interval of $\psi$. Thus, $\psi$ assumes a maximum when $\phaselag=\pi/2$. It can be inferred from \eref{rhotrafo} that this corresponds to a maximum localization in $a_2$. In other words, at the point of pure out-of-unison motion with a $90^\circ$ phase lag between the modes, the motion localizes in the second mode to a locally maximum extent.\\
For the point of maximum mode complexity, the associated value of the normalized detuning can be identified from \eref{sincosdelta}. In can be expressed as a function of $\cos\psi$,
\e{\sigma^*_{\frac{\pi}{2}} = \frac{1+\cos^2\psi}{4} \in \left[\frac14\,,\,\,\frac12\right]\fp}{sigst}
This defines the normalized detuning range in which the point of maximum mode complexity can be encountered. For small damping, the value will be close to $1/4$, while it will approach $1/2$ for large damping.

\section{Illustration of the results\label{sec:results}}
In this section, the analytical results derived in \sref{msa} are illustrated and validated, and some insights into the dynamic behavior beyond the analytically investigated regime are given. \red{In \ssref{detuning}, the typical dependence of mode complexity and localization on detuning is presented. The influence of damping is investigated in \ssref{damping}. A comparison of the forced response with the corresponding nonlinear normal modes is addressed in \ssref{undampedFRF_vs_NNM}. A validation of the asymptotic analysis against an accurate reference is given in \ssref{validate}. Finally, some insight into the quasi-periodic vibration regime is provided in \ssref{qperiodic}.}

\subsection{Typical dependence of the response on detuning and transition from standing to traveling waves\label{sec:detuning}}
\figw[t]{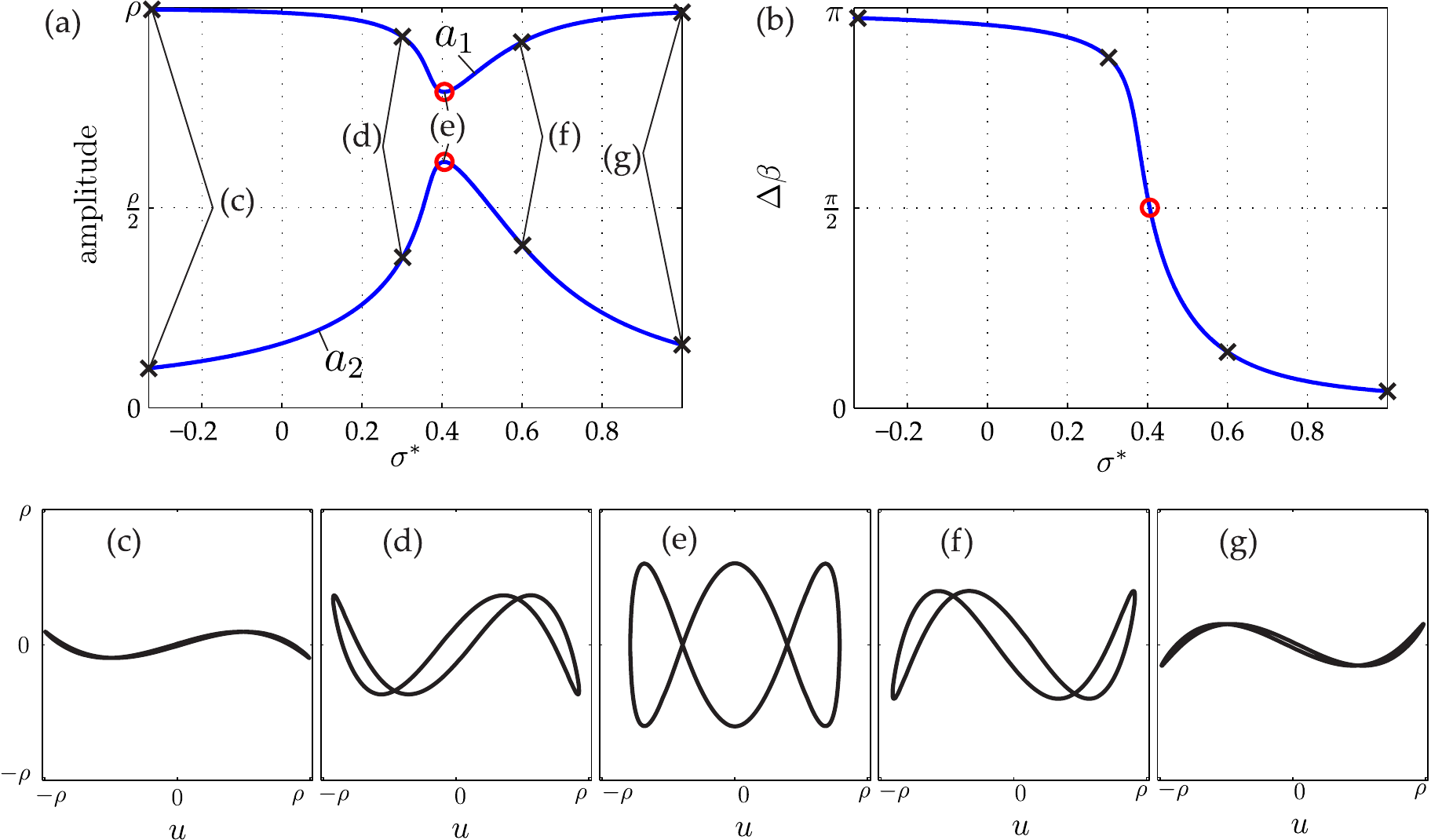}{Typical dependence of the response on the detuning parameter $\sigma^*$: (a) modal amplitudes, (b) phase lag between the modes, (c)-(g) periodic motions illustrated in the modal coordinate space; the red circle indicates the point of maximum mode complexity}
For a specific value of the normalized damping parameter $\del^*=0.067$, the dependence of the response on the detuning parameter is shown in \fref{typical_dst}. As was concluded from the analysis, the second modal amplitude $a_2$ attains a local maximum, in this case at $\sigma^*=0.4$, which is in the predicted range between $1/4$ and $1/2$, \cf \eref{sigst}. This coincides with the location of a minimum of the first modal amplitude $a_1$, and a phase lag of $\phaselag=\pi/2$. The associated phase projection is illustrated in \fref{typical_dst}e. Apparently a large area is enclosed in this Lissajous curve. Away from this critical point, the motion localizes in the first mode, and a only a small area is enclosed in the projected orbit, \cf \fref{typical_dst}c-d and f-g.
\figc[t]{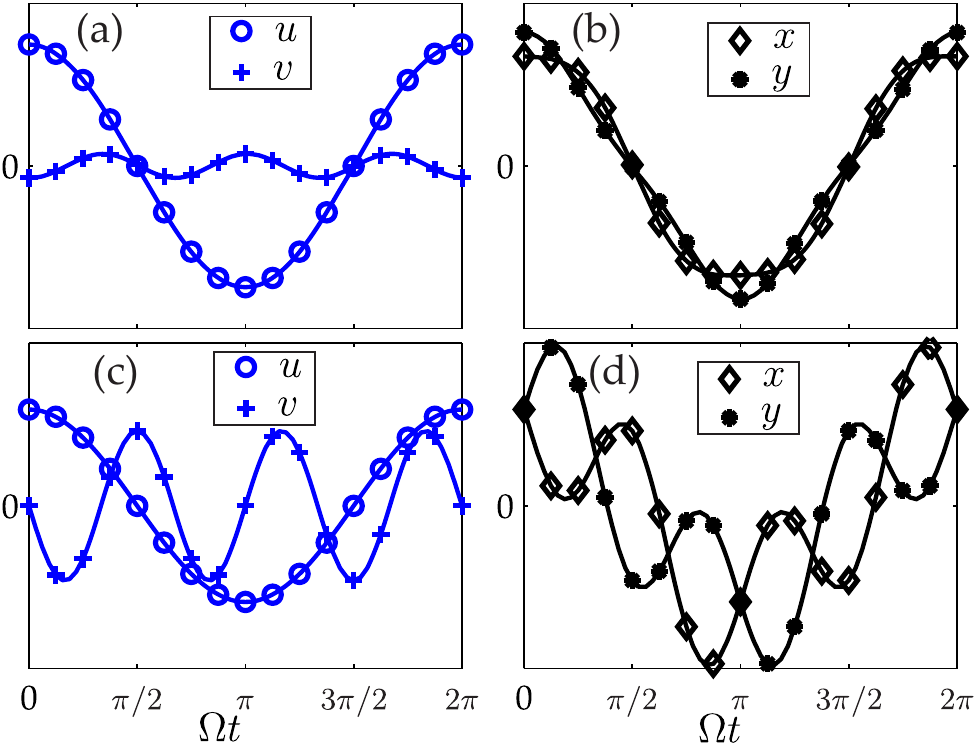}{Time history of the steady-state response in modal coordinates $u,v$ and physical coordinates $x,y$, \red{$\del^*=0.067$}: (a)-(b) synchronous motion away from the internal resonance, $\sigma^*=-0.3$; (c)-(d) anti-synchronous motion in the neighborhood of the internal resonance, $\sigma^*=\sigma^*_{\frac{\pi}{2}}$}
\\
The essential transition from synchronous to anti-synchronous motion can also be inferred from the time histories depicted in \fref{typical_dst_SW_TW}. Away from resonance, the modes oscillate synchronously in a multi-harmonic sense; \ie zero-crossings and turning points of the low-frequency motion $u(t)$ coincide with zero-crossings and turning points of the fast-frequency motion $v(t)$, \cf \fref{typical_dst_SW_TW}a. At the internal resonance, in contrast, the modes oscillate anti-synchronously. Consequently, zero-crossings of $u(t)$ coincide with turning points of $v(t)$, and turning points of $u(t)$ coincide with zero-crossings of $v(t)$, \cf \fref{typical_dst_SW_TW}c. In a generalized sense, the transition from synchronous to anti-synchronous motion can be interpreted as a \textit{transition from standing to traveling waves}. It should be noted that this finding only applies to the dynamic behavior in the modal space, while the physical coordinates only oscillate in a general non-synchronous fashion, \cf \fref{typical_dst_SW_TW}b,d.

\subsection{Dependence of mode complexity and localization on damping and detuning\label{sec:damping}}
\figw[t]{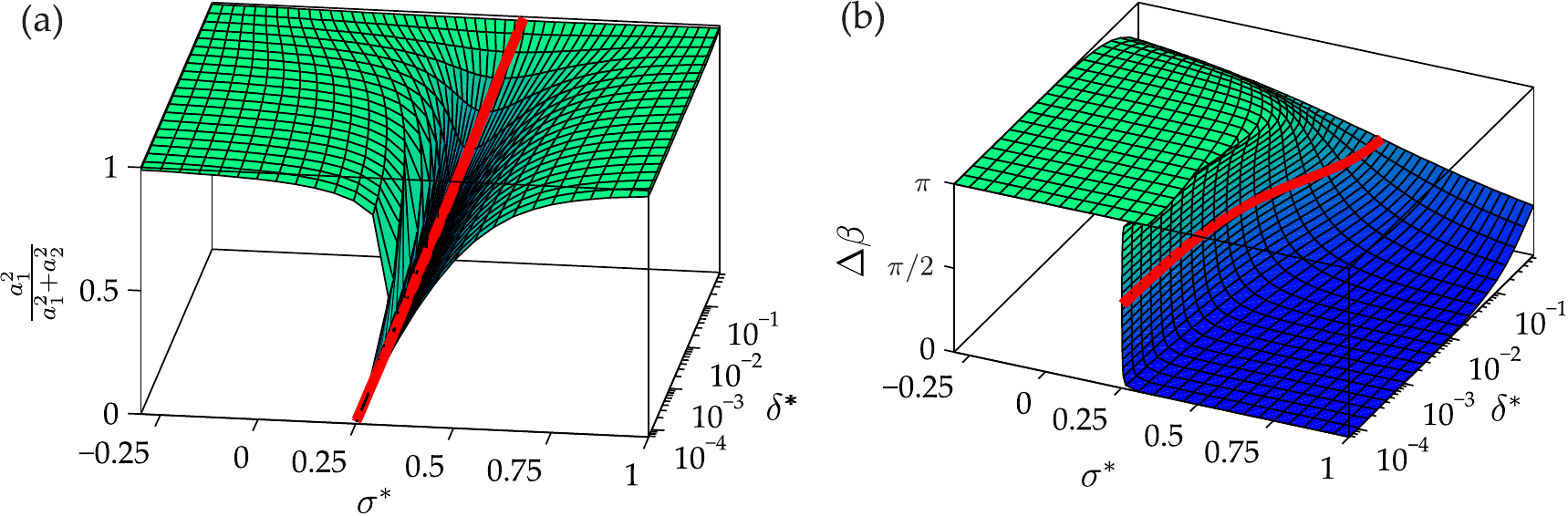}{Dependence of the response on the detuning parameter $\sigma^*$ and the damping parameter $\del^*$: (a) localization in $a_1$, (b) phase lag between the modes; the red curve indicates the curve of maximum mode complexity}
Next, the dependence of the normalized damping parameter $\del^*$ is illustrated in addition to the detuning. To this end, a measure for the localization in $a_1$ is introduced,
\e{\frac{a_1^2}{a_1^2+a_2^2}\fp}{localization_measure}
This quantity tends towards unity if $a_1\gg a_2$, it tends towards zero if $a_1\ll a_2$, and it assumes the value $1/2$ if $a_1=a_2$. In \fref{loc_delta_vs_dst_sigst}, the localization in $a_1$ and the mode complexity, measured by the phase lag $\phaselag$, are depicted.\\
When $\del^*\rightarrow 0$, the motion localizes completely in $a_2$ at the super-harmonic resonance, and therefore the localization in $a_1$ \erefo{localization_measure} tends towards zero, \cf \fref{loc_delta_vs_dst_sigst}a. In contrast to \fref{typical_dst}, there are ranges with multiple solutions. Only two of these motions are asymptotically stable. The aspect of stability will be further investigated in \sref{validate}.\\
For increasing $\del^*$, the super-harmonic resonance is damped away. Moreover, the phase lag characteristic flattens out for increasing damping. The super-harmonic resonance is clearly bounded between $1/4$ and $1/2$, in full accordance with \eref{sigst}.

\subsection{Comparison of the undamped forced response with the nonlinear normal mode\label{sec:undampedFRF_vs_NNM}}
\fig[t]{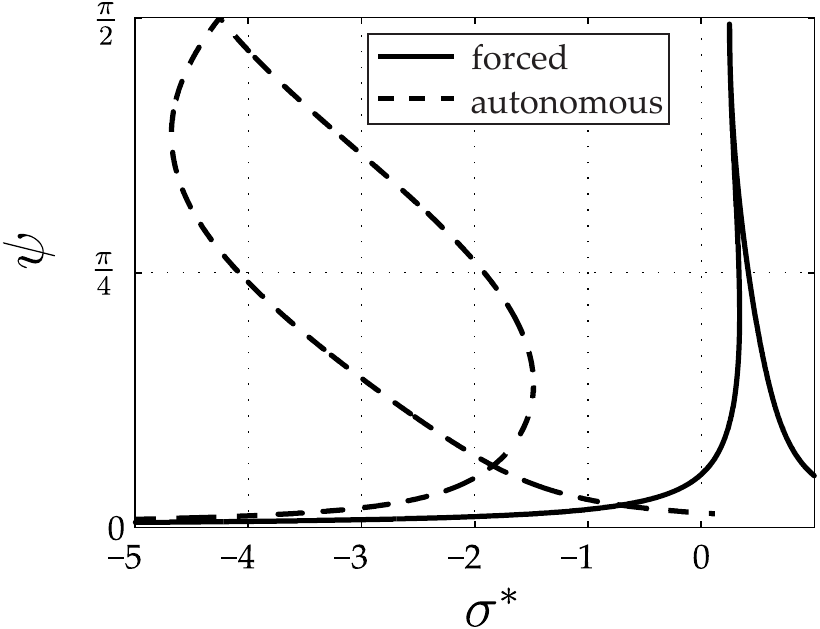}{Comparison of the modal interaction in the undamped forced response and in the nonlinear normal mode}
In an earlier study, the considered two-degree-of-freedom system was investigated in its autonomous and undamped setting \cite{vaka2008}. In the neighborhood of the $1:3$ internal resonance, the only existing non-similar nonlinear normal mode was analyzed. The corresponding slow flow equations can be easily obtained from \erefs{slowflowa}-\erefo{slowflow} by setting the forcing amplitude, external detuning and damping to zero. It was demonstrated that, in this case, the quantity $a_1^2+9a_2^2$ is conserved and the dynamics take place on an iso-energetic two-torus. We will focus on the periodic solutions in the following. With the coordinate transformation \erefo{rhotrafo} and the definition of $\sigma^*$ used in the present work, the equation governing $\sin\psi$ again takes a bi-cubic form
\ea{\nonumber &1684{\sin^6\psi} + \left(720\sigma^*+420\right){\sin^4\psi} + \\
&\quad\quad\quad\left(144{\sigma^*}^2+504{\sigma^*}+498\right){\sin^2\psi} = 1\fp}{bicubic_autonomous}
The results are compared in \fref{autonomous_vs_forced}. It should be noted that $\sigma^*$ defined in \eref{sincosdelta} depends on system parameters as well as the vibration level $\rho$. Although the interesting phenomena take place in different ranges of this parameter, the qualitative behavior is indeed quite similar. Far away from the tuned case, \ie, $\left|\sigma^*\right|\gg 0$, $\psi$ tends towards zero, and thus the motion localizes in the first mode $a_1$, \cf \eref{rhotrafo}. At a certain detuning value, $\psi$ reaches a maximum of $\pi/2$, which corresponds to localization in $a_2$.

\subsection{Validation of the approximation\label{sec:validate}}
The analytical multiple scales approximation was validated using numerical time integration \red{of the initial equations of motion \erefo{eqm_uva}-\erefo{eqm_uv}}. To directly compute the periodic steady-state oscillations, the well-known shooting method was applied. The associated boundary value problem was solved with a Newton-like method in conjunction with an arc-length predictor-corrector path continuation technique. The constant average acceleration Newmark scheme was utilized for time integration; $1024$ time steps per excitation period were deemed sufficient according to a preliminary convergence study. The Jacobian matrix of the boundary value problem was obtained by integrating the analytical tangent stiffness matrix. Thus, the monodromy matrix was a by-product of the computational procedure, and its eigenvalues were computed for classical Floquet stability analysis. In \fref{frf_validation}, the approximated modal amplitudes and the phase lag are compared to the numerical reference for the parameter set listed in the figure caption. In the case of the shooting method, $a_1$ and $a_2$ are defined as the first harmonic of $u(t)$ and the third harmonic of $v(t)$, respectively, and can be easily computed by applying the Fourier transform to the periodic solutions.\\
In the range around $\omone=1$, for which the multiple-scales analysis has validity, the asymptotic results are in very good agreement with the numerical reference. For larger excitation frequencies, however, the approximation is poor, indicating that higher-order effects become important. The first mode reaches a comparatively large amplitude in the neighborhood of its fundamental resonance here around $\Omega=1.8$. 
Due to the considerable stiffening effect, the fundamental resonance occurs at a frequency which is already much larger than its linear counterpart $\omone=1$.
\figw[t]{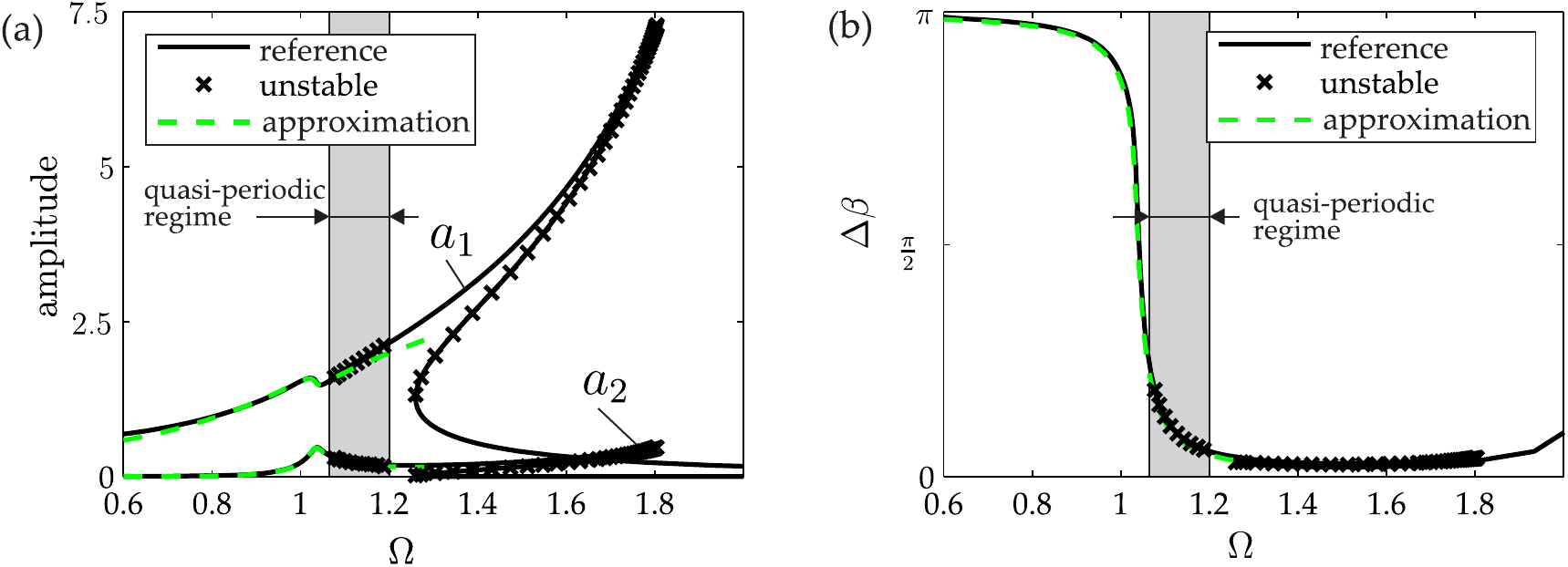}{Frequency response function: (a) amplitudes, (b) phase lag; $\omone=1$, $\omtwo=3$, $\del=0.06$, $\gam=0.1$, $\fuh=0.5=\fvh$}
\\
In the frequency ranges where multiple steady-state responses co-exist, saddle-node bifurcations take place at the turning points, and the overhanging branch is unstable. In addition, Torus bifurcations\footnote{`Secondary Hopf bifurcation' and `Neimark-Sacker bifurcation' are also widely-used terms in this context.} give rise to a quasi-periodic regime as indicated in \fref{frf_validation}. This regime will be investigated in more detail in the next subsection. In the strongly nonlinear regimes, more bifurcations may occur, and isolated solution branches can generally be expected; see \eg \cite{grolet2014}.

\subsection{The quasi-periodic regime\label{sec:qperiodic}}
Quasi-periodic motions are commonly encountered in the forced response of nonlinear systems. To provide further insight into the frequency content of the dynamics, the response spectra of the first and the second modal coordinates are illustrated in \fref{qp_spectrum_vs_Om}a and b, respectively. They have been obtained by first simulating \eref{eqm_uva}-\erefo{eqm_uv} from homogeneous initial conditions for $2000$ excitation periods. Then the Fourier transform was applied to the last $1000$ excitation periods which were assumed to reflect the steady-state behavior. A Hanning window was utilized for the analysis of the non-periodic time signal. Although $u(t)$ also possessed higher harmonics, it was found that its spectrum was still dominated by its fundamental harmonic component. Similarly, $v(t)$ was dominated by the third harmonic. Hence, the spectra in \fref{qp_spectrum_vs_Om} are only shown for the relevant range around their dominant frequencies.\\
A second frequency $\tilde\omega$ is generated in the quasi-periodic regime, where $\tilde\omega$ and $\Omega$ are incommensurable. A time signal comprising these two base frequencies can in general be written as
\e{x(t) = \real{\sum\limits_{n}^{}\sum\limits_{m}^{}\hat x_{nm}\ee^{\ii\left(n\Omega t+m\tilde\omega t\right)}\fk}}{qp_general}
where $n$, $m$ are integers, and $\hat x_{nm}$ are complex amplitudes. As stated above, the fundamental harmonic $n=1$ of the excitation frequency dominates the response of $u(t)$, while the third harmonic $n=3$ dominates $v(t)$. \eref{qp_general} can then be simplified as follows:
\e{u(t) \approx \real{\underbrace{\left[\sum\limits_{m}^{}\hat u_{1m}\ee^{\ii m\tilde\omega t}\right]}_{2a_1(\tilde\omega t)~\ee^{\ii\beta_1(\tilde\omega t)}}\ee^{\ii\Omega t}}\fp}{qp_simplified}
This corresponds to the form of \eref{uv_solution}; now, however, the amplitude $a_1$ and the phase lag $\beta_1$ are periodic functions with fundamental frequency $\tilde\omega$. The transformation in \eref{qp_simplified} is the key to the generalization of the notion of amplitude and phase lag, and thus localization and mode complexity, to quasi-periodic motions.
\figw[t]{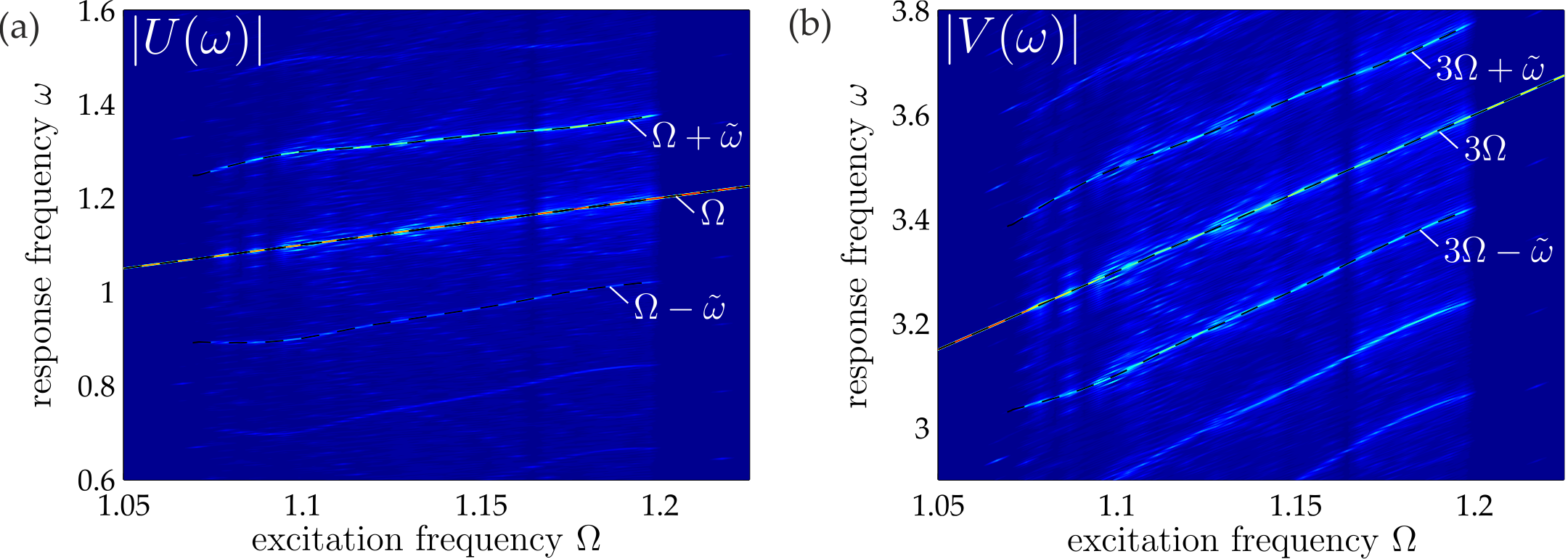}{Response frequency spectrum in the vicinity of the quasi-periodic regime: (a) spectrum of the first modal coordinate $u(t)$, (b) spectrum of the second modal coordinate $v(t)$}
\figw[t]{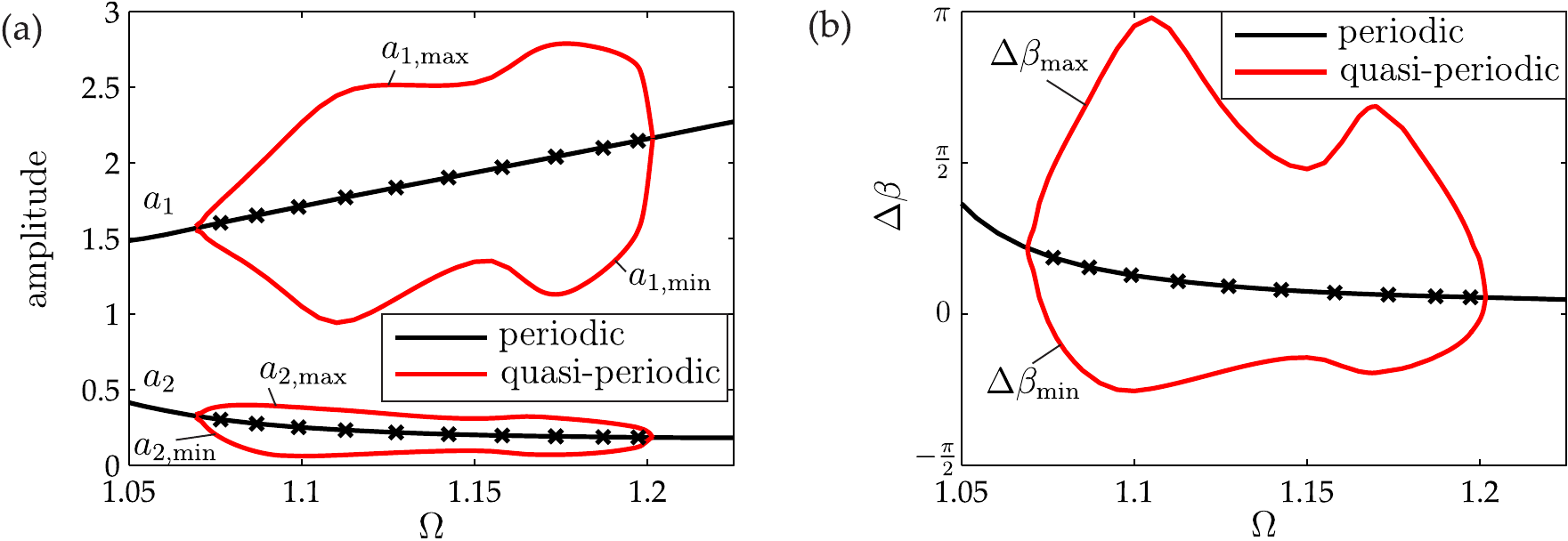}{Bifurcation diagram of the forced response in the vicinity of the quasi-periodic regime: (a) amplitudes, (b) phase lag; crosses indicate the unstable periodic regime}
\\
It can be clearly seen from \fref{qp_spectrum_vs_Om} that mainly $\Omega$ ($3\Omega$) occurs in the spectrum of $u$ ($v$) outside the quasi-periodic region, \ie, for $\Omega<1.075$ and $\Omega>1.2$. Within the quasi-periodic region, the frequencies $\Omega\pm\tilde\omega$ ($3\Omega\pm\tilde\omega$) are also present, in full accordance with \erefs{qp_general} and \erefo{qp_simplified}. Apparently, higher integers $\left|m\right|>1$ are also important in the spectrum of $v$.\\
The quasi-periodic attractor was also directly computed using a \red{Fourier method \cite{schilder2006}}. To this end, the ansatz \erefo{qp_general} was considered with $m$ and $n$ ranging from $-7$ to $+7$. To improve numerical performance, redundant pairs of harmonic indices were removed. In addition to the complex amplitudes $\hat u_{nm}$ and $\hat v_{nm}$, the second frequency $\tilde\omega$ was also treated as an unknown. The problem was solved in the frequency domain and involved two-dimensional variants of the discrete Fourier transform to determine the complex amplitudes related to the nonlinear forces. Using a continuation procedure with respect to the excitation frequency, the computation was carried out in the whole quasi-periodic regime. For each solution point, the periodic amplitudes $a_1$, $a_2$ and the phase lag $\phaselag$ were determined. Maximum and minimum values of these quantities are indicated in \fref{qp_bifurcation_diagram} in addition to their constant counterparts in the periodic case. The maximum values of the amplitudes and the phase lag are always larger than their corresponding values in the unstable periodic case, while their minimum values are always smaller than in the periodic case. Within the quasi-periodic regime, the oscillatory part is in the order of magnitude of the static component. As expected, the oscillatory part tends towards zero when approaching the bifurcation points from the quasi-periodic regime.
\figw[t]{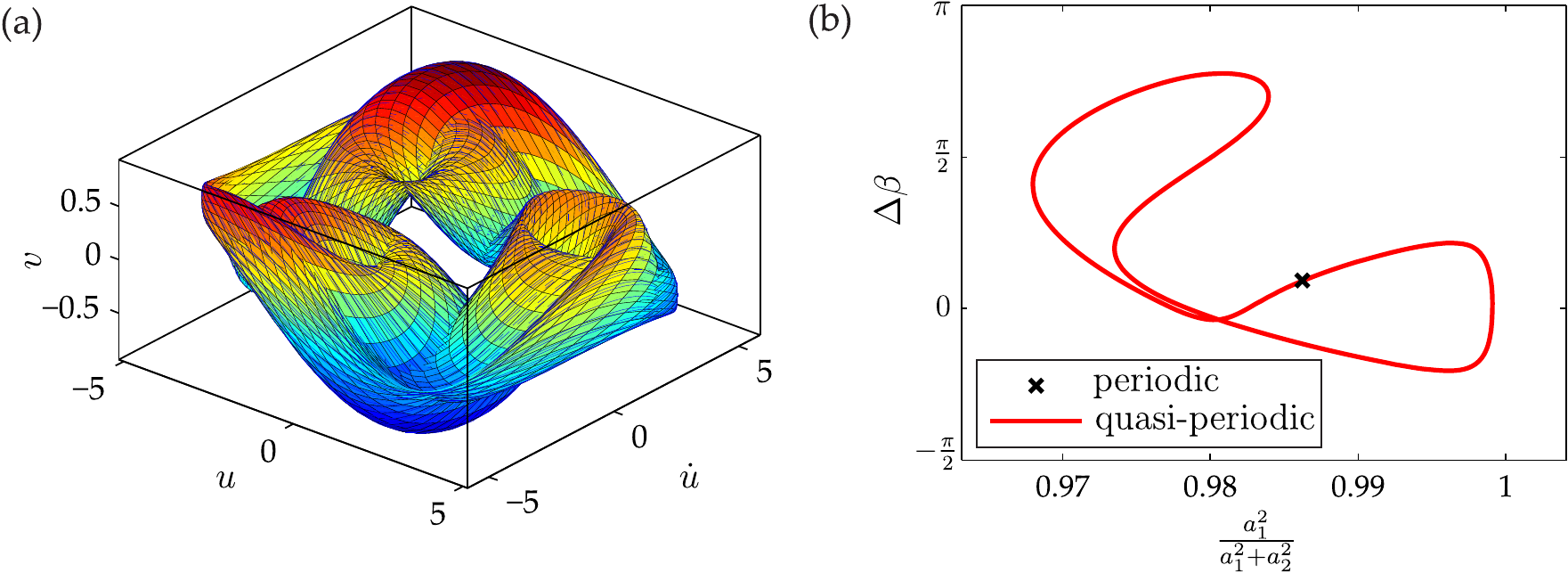}{Dynamic behavior for $\Omega=1.12$: (a) simulated trajectory (in blue) covering the invariant torus, (b) periodically oscillating phase lag and localization in $a_1$}
\\
For $\Omega=1.12$, the quasi-periodic motion is illustrated in \fref{qp_torus_and_slice}. The trajectory simulated with numerical time integration covers the invariant torus computed with the Fourier method. The localization in the first mode and the mode complexity are also shown in \fref{qp_torus_and_slice}b. Since the amplitude $a_2$ of the second mode is generally smaller than the amplitude $a_1$ of the first mode, the localization in the first mode assumes values close to unity. While the periodic motion corresponds to a fixed point, mode complexity and localization oscillate with a frequency $\tilde\omega$ in the the quasi-periodic case. The closed curve assumes the form of a stretched and folded $8$, and thus both mode complexity and localization reach two local minima and two local maxima within one period of the second frequency.

\section{Conclusions\label{sec:conclusions}}
In contrast to the common assumption that weak, linear damping and resonant forcing merely affect the level of vibration, we demonstrated that they may indeed globally change the qualitative nature of the vibration. This global effect is mode complexity, \ie, asynchronous modal oscillations in the neighborhood of an internal resonance. It was generally ascertained that mode complexity and localization strongly depend on damping and detuning. The analytical study revealed that a point of maximum possible phase lag always occurs and is located within a bounded range of the detuning parameter. This coincides with the location of the super-harmonic resonance of the indirectly excited mode, associated with a maximum of the localization in this mode. In the absence of damping, this localization in the indirectly excited mode develops to a full extent, while the motion localizes in the directly excited mode for large damping. Hence, there exists a specific damping value for which the mode mixing is most prominent and the modes oscillate with a phase lag of $90^\circ$ in a multi-harmonic sense. This anti-synchronous motion can be interpreted as a generalized traveling wave in the system's modal space. In this sense, a transition from standing to traveling waves takes place in the neighborhood of the internal resonance.\\
Moreover, the torus bifurcation giving rise to quasi-periodic motions in the neighborhood of the resonance coincidence was numerically investigated. In the considered weakly nonlinear regime, the frequency content of the modes is still dominated by their respective first and third harmonics of the excitation frequency. \red{As demonstrated}, this permits the generalization of the notion of amplitude and phase lag, and thus mode localization and mode complexity, to quasi-periodic motions. To the authors' knowledge, such a generalization has not yet been proposed in the literature. Amplitudes and phase lags featured a strong oscillatory part as compared to their static component.\\
It is conjectured that the findings for the specific two-degree-of-freedom system can be extended to generic systems that are close to a $1:3$ internal resonance condition. It is further assumed that the knowledge gained through this work provides an apt basis for the development of model order reduction techniques. The purpose of such techniques would be the reduction of systems described by an arbitrary number of degrees of freedom down to only the coordinates of the participating nonlinear modes.

\begin{acknowledgements}
The work leading to this publication was supported by the German Academic Exchange Service (DAAD) with funds from the German Federal Ministry of Education and Research (BMBF) and the People Programme (Marie Curie Actions) of the European Unions Seventh Framework Programme (FP7/2007-2013) under REA grant agreement no. 605728 (P.R.I.M.E. - Postdoctoral Researchers International Mobility Experience).
\end{acknowledgements}

\section*{Compliance with Ethical Standards}
Funding: This study was funded by the German Academic Exchange Service within the P.R.I.M.E. program, see acknowledgements.\\
Conflict of Interest: The authors declare that they have no conflict of interest.

\begin{appendix}
\section{Derivation of the slow flow equations\label{asec:derivation_slowflow}}
In this appendix, we describe the important mathematical steps involved in the derivation of \erefs{slowflowa}-\erefo{slowflow}. As usual, we expand the solution $u,v$ in a perturbation series,
\eal{
u\left(\epsilon,t\right) &= u_0\left(\tsclone,\tscltwo\right) + \epsilon u_1\left(\tsclone,\tscltwo\right)+\order{\epsilon^2}\fk\\
v\left(\epsilon,t\right) &= v_0\left(\tsclone,\tscltwo\right) + \epsilon v_1\left(\tsclone,\tscltwo\right)+\order{\epsilon^2}\fk}{perturbation_series}
with the fast time scale $\tsclone$ and the slow time scale $\tscltwo$ given by,
\e{\tsclone = t \fk\quad \tscltwo = \epsilon t\fp}{timescales}
The time derivative operator becomes
\e{\frac{\dd}{\dd t} = \frac{\partial}{\partial\tsclone} + \epsilon\frac{\partial}{\partial\tscltwo}\fp}{timederivative}
\erefs{perturbation_series}-\erefo{timederivative} are then substituted into the rescaled version of \erefs{eqm_uva}-\erefo{eqm_uv}. Balancing the $\order{1}$ terms yields,
\ea{\frac{\partial^2 u_0}{\partial\tsclone^2}& + \omonesq u_0& = 0\fk\label{eq:orderonea}\\
\frac{\partial^2 v_0}{\partial\tsclone^2}& + 9\omonesq v_0& = 0\fp}{orderone}
The corresponding $\order{\epsilon}$ equations read as follows:
\ea{\frac{\partial^2 u_1}{\partial\tsclone^2}& + \omonesq u_1 &= -2\frac{\partial^2 u_0}{\partial\tsclone\partial\tscltwo} - \frac \gam 2\left(u_0+v_0\right)^3 + f_u \fk\label{eq:ordertwoa}\\
\frac{\partial^2 v_1}{\partial\tsclone^2}& + 9\omonesq v_1 &= -2\frac{\partial^2 v_0}{\partial\tsclone\partial\tscltwo} - \frac \gam 2\left(u_0+v_0\right)^3 + f_v \nonumber \\
&& \quad\quad - 2\del\frac{\partial v_0}{\partial\tsclone} - 6\omone\sigtwo v_0\fp}{ordertwo}
The general solution of \erefs{orderonea}-\erefo{orderone} is $u_0 = A(\tscltwo)\ee^{\ii\omone\tsclone}+\cc$ and $v_0 = B(\tscltwo)\ee^{3\ii\omone\tsclone}+\cc$. Substituting this into \erefs{ordertwoa}-\erefo{ordertwo}, and requiring the secular terms to vanish, yields
\ea{
0= -2\ii\omone A^\prime & - \frac{3\gam}{2}\left[A^2\conj{A}+\conj{A}^2B+2B\conj{B}A\right]&+\fuh\ee^{\ii\sigone \tscltwo}\fk \label{eq:slowflow_complexa}\\
0= -6\ii\omone B^\prime & - \frac{\gam}{2}\left[A^3+3B^2\conj{B}+6A\conj{A}B\right]& \nonumber\\
 & -6\omone\left(\sigtwo+\ii \del\right)B \fp
}{slowflow_complex}
Herein, ${}^\prime$ denotes derivative with respect to $\tscltwo$. It should be noted that the term \g{\fvh} does not occur in these equations. Therefore a weak fundamental harmonic forcing of the out-of-phase mode has only second-order effects. Without loss of generality of the subsequent investigations, we assume \g{\fuh=\fone>0} as a positive real-valued quantity in the following.\\
We introduce polar coordinates \g{A=a_1\ee^{\ii\left(\sigone\tscltwo+\beta_1\right)}} and \g{B=a_2\ee^{3\ii\left(\sigone\tscltwo+\beta_2\right)}} with the real-valued quantities $a_1$, $a_2$, $\beta_1$ and $\beta_2$. Substitution into \erefs{slowflow_complexa}-\erefo{slowflow_complex} and separation of real and imaginary part finally gives rise to the slow flow equations \erefo{slowflowa}-\erefo{slowflow}.
\end{appendix}


\end{document}